\documentclass[12pt]{article}

\usepackage{amsfonts}

\newcommand{\bi}{\bibitem}

\begin{document}
\title{The Dirac field in Taub-NUT background}

\author{Ion I. Cot\u aescu \thanks{E-mail:~~~cota@quasar.physics.uvt.ro}\\ 
{\small \it The West University of Timi\c soara,}\\
       {\small \it V. P\^ arvan Ave. 4, RO-1900 Timi\c soara, Romania}
\and
Mihai Visinescu \thanks{E-mail:~~~mvisin@theor1.theory.nipne.ro}\\
{\small \it Department of Theoretical Physics,}\\
{\small \it National Institute for Physics and Nuclear Engineering,}\\
{\small \it P.O.Box M.G.-6, Magurele, Bucharest, Romania}}
\date{\today}

\maketitle

\begin{abstract}
We investigate the $SO(4,1)$ gauge-invariant theory of the Dirac fermions 
in the external field of the Kaluza-Klein monopole, pointing out that the 
quantum modes can be recovered from a Klein-Gordon equation analogous to the 
Schr\" odinger equation in the Taub-NUT background. Moreover, we show that 
there is a large collection of observables  that can be directly derived 
from those of the scalar theory. These offer many possibilities of choosing 
complete sets of commuting operators which determine the quantum modes. 
In addition there are some spin-like and Dirac-type operators involving 
the covariantly constant Killing-Yano tensors of the hyper-K\"ahler 
Taub-NUT space. The energy eigenspinors of the central modes in spherical 
coordinates are completely evaluated in explicit, closed form. 

Pacs 04..62.+v

\end{abstract}

\newpage

\section{Introduction}
\

Hawking \cite{Ha} has suggested that the Euclidean Taub-NUT metric 
might give rise to the gravitational analogue of the Yang-Mills 
instanton. The Euclidean Taub-NUT metric satisfies Einstein's equations 
with zero cosmological constant. 
This  metric  is also involved in many modern studies in physics.
For example the Taub-NUT metric is the space part of the line element 
of the celebrated Kaluza-Klein monopole of Gross and Perry \cite{GP} and 
of Sorkin \cite{So}. On the other hand, in the 
long-distance limit, neglecting radiation, the relative motion of two 
monopoles is described by the geodesics of this space \cite{G1}.
From the mathematical point of view, the Taub-NUT geometry is also very 
interesting. In the Taub-NUT geometry there are four Killing-Yano tensors 
\cite{GR}. Three of these are complex structure realizing the quaternionic 
algebra and the Taub-NUT manifold is hyper-K\" ahler. In addition to these 
three vector-like Killing-Yano tensors, there is a scalar one and it 
exists by virtue of the metric being type $D$.

The Schr\" odinger quantum modes in the Euclidean Taub-NUT geometry were 
analyzed  using algebraic \cite{G2} or analytical methods \cite{CV}.
The Dirac equation in this background was studied in the mid eighties 
\cite{DIRAC}. It was later realized that the geodesic motion in 
Euclidean Taub-NUT space is integrable and has a remarkable close 
similarity with motion under a Coulomb force \cite{GR,G2}. For the 
geodesic motion in the Taub-NUT space, the conserved vector analogous 
to the Runge-Lenz vector of the Kepler type problem is quadratic in 
$4$-velocities, its components are St\" ackel-Killing tensors and they 
can be expressed as symmetrized product of Killing-Yano tensors.

The fermion problem in the Taub-NUT gravitational instanton field was 
also studied by Comtet and Horvathy \cite{CH} using the observation 
that the Dirac operator is supersymmetric in $4$ dimensions. This 
property can be connected with fact that the Pauli Hamiltonian for a 
spin $1\over 2$ particle in the field of a Dirac magnetic monopole 
possesses a dynamical supersymmetry \cite{DYON}. 

We should like to continue this study in the context of the standard 
relativistic gauge-invariant theory \cite{W,BD} of the Dirac field in 
Taub-NUT background \cite{DIRAC}. There are some gaps in the previous 
treatments of the problem which must be filled in. For example, in the 
Schr\" odinger case, the existence of the extra conserved quantities of 
the Runge-Lenz type implies the possibility of separating variables in 
two different coordinate systems. The presence of the "hidden" symmetry 
of the Taub-NUT problem must exercise an influence over the Dirac 
equation in this background. Finally, we intend to compare the study of 
the Dirac equation on the Taub-NUT manifold with other treatments of 
the fermions on curved spaces. For example, there are pseudo-classical 
models for relativistic spin ${1\over 2}$ particle involving anticommuting 
vectorial degrees of freedom \cite{grass}. For the Taub-NUT spinning 
space, the relations between symmetries, supersymmetries and constants 
of motion have been investigated in \cite{G3}. Spinning particles are in 
some sense the classical limit of the Dirac particles. The quantization of 
these models gives rise to supersymmetric quantum mechanics. After 
quantization, the anticommuting Grassmann variables are mapped into the 
Dirac matrices while the  supercharge becomes just the static part of 
the Dirac operator (as defined in section 3).

We devote the present paper to the general $SO(4,1)$ gauge-invariant theory 
of the Dirac fermions \cite{DKK} in the external field of the Kaluza-Klein 
monopole\cite{DIRAC}. Our goal is to point out new  features of this theory 
and to find  the quantum modes determined by complete sets of commuting 
operators. We start with a gauge-invariant action involving the Dirac field 
and {\em pentad} fields (or f\" unfbein) giving local frames like in 
Ref.\cite{P}. Thus we  preserve the manifest covariance of the theory under 
space rotations such that the  familiar three-dimensional vector notation can 
be used. Our main result is that the Dirac equation obtained in this way 
\cite{DIRAC} is analytically solvable having quantum modes that can be 
recovered from those of the Klein-Gordon equation which has the same solutions 
as the Schr\" odinger equation \cite{CV} but with different parameters. Thus 
we show that the Dirac equation can be solved in a manner close to that used 
for deriving the eigenspinors of the static Dirac operator studied in 
Ref.\cite{CH}. Another result is that for each conserved observable of the 
scalar theory there is  at least one operator which commutes with the Dirac 
Hamiltonian. Consequently, new conserved observables arise including some 
interesting ones that can be related to the specific objects of the Taub-NUT 
geometry. As an application, we write down the general form of the energy 
eigenspinors of the central modes determined by a complete set of commuting 
operators that includes the Hamiltonian. We obtain a similar energy spectrum 
as in Ref.\cite{CH} but different energy eigenspinors written in terms of new 
spherical spinors which solve the angular eigenvalue problems. 

We start in section 2 with a brief review of the Taub-NUT geometry and the 
main orbital operators. In the next one we derive the Dirac equation from 
the standard $SO(4,1)$ gauge-invariant action  while in section 4 we show 
that this theory has well defined properties  concerning  the supersymmetry 
and Hermitian conjugation. In section 5 we  solve the Dirac equation while 
in the next one we discuss the form of the possible conserved observables. 
Section 7 is devoted to the construction of spin-like and Dirac-type 
operators in direct connections with the hyper-K\" ahler structure of the 
Euclidean Taub-NUT space. In  section 8  we give the solutions corresponding 
to the central modes and, finally, we present our conclusions. In Appendix 
we introduce the new spherical spinors we need.  We work in natural units 
with $\hbar=c=1$.

\section{Preliminaries}
\

The Kaluza-Klein monopole \cite{GP,So} was obtained by embedding the 
Taub-NUT gravitational instanton into five-dimensional theory, adding 
the time coordinate in a trivial way. In a static chart of 
coordinates $x^{\mu}$ ($\mu, \nu,...=0,1,2,3,5$), its line element is 
expressed as
\begin{equation}\label{(met)} 
ds^{2}=g_{\mu\nu}dx^{\mu}dx^{\nu}=dt^{2}-\frac{1}{V}dl^{2}-V(dx^{5}+
A_{i}dx^{i})^{2}\,.
\end{equation}   
Here $dl^{2}=(d\vec{x})^{2}=(dx^{1})^{2}+(dx^{2})^{2}+(dx^{3})^{2}$
is the usual Euclidean 3-dimensional line element involving the Cartesian 
physical space coordinates $x^{i}$ ($i,j,...=1,2,3$) which 
cover the domain $D$. The other coordinates are the time, $x^{0}=t$, and 
the Cartesian Kaluza-Klein extra-coordinate, $x^{5}\in D_{5}$. 
The functions  $V$ and $A_{i}$ are static  depending only on $\vec{x}$ as     
\begin{equation}\label{(tn)}
\frac{1}{V}=1+\frac{\mu}{r}\,,\quad A_{1}=-\frac{\mu}{r}\frac{x^{2}}
{r+x^{3}}\,,\quad 
A_{2}=\frac{\mu}{r}\frac{x^{1}}{r+x^{3}}\,,\quad A_{3}=0 
\end{equation}
where $r=|\vec{x}|$. The regular Taub-NUT metric has the function $V$ 
with $\mu$ positive and it is smooth in the range $r\geq 0$.  
The case of $\mu<0$ is also interesting since then scalar modes with discrete 
energy levels are allowed. In both cases  we consider that the space 
domain of the local chart with Cartesian coordinates is defined by the 
condition $V>0$.  ${\vec A}$ 
is the Dirac monopole vector potential giving the  magnetic field 
with central symmetry
\begin{equation}
\vec{B}\,={\rm rot}\, \vec{A}=\mu\frac{\vec{x}}{r^3}\,.
\end{equation}
In fact,  the  spacetime defined by (\ref{(met)}) and (\ref{(tn)}) has the 
{\em global} symmetry 
of the  group $G_{s}=SO(3)\otimes U_{5}(1)\otimes T_{t}(1)$ since  
the line element is invariant under the global rotations of the Cartesian space 
coordinates and  $x^{5}$ and $t$ translations of the Abelian groups 
$U_{5}(1)$ and $T_{t}(1)$ respectively. We note that the $U_{5}(1)$ symmetry 
eliminates the so called NUT singularity  if $x^5$ has the period 
$4\pi\mu$.

The main orbital operators of the relativistic quantum mechanics  can be 
introduced by using the geometric quantization. In this way, one obtains  
the  momentum operators in coordinate representation, 
\begin{equation}
P_{i}=-i(\partial_{i}-A_{i}\partial_{5})\,,\quad P_{5}=-i\partial_{5}\,.
\end{equation} 
They obey the commutation rules
\begin{equation}
[P_{i},P_{j}]=i\varepsilon_{ijk}B_{k}P_{5}\,,\quad
[P_{i},P_{5}]=0\,,
\end{equation}
and give the Klein-Gordon operator,   
\begin{equation}
\nabla_{\mu}g^{\mu\nu}\nabla_{\nu}=
{\partial_{t}}^{2}+\Delta\,,\quad
\Delta= V{\vec{P}\,}^{2}+\frac{1}{V}{P_{5}}^{2}\,,
\end{equation}
where $\nabla_{\mu}$ are the usual covariant derivatives.

Other important orbital operators are the generators of the group $G_{s}$ 
defined up to the factor $-i$ as  the Killing vector fields corresponding to 
the global symmetry of the background. Thus the generator of the group 
$U_{t}(1)$ is $-i\partial_{t}$ while the $U_{5}(1)$ generator is just $P_{5}$.  
The other three Killing vectors give the $SO(3)$ generators 
which are the components of the orbital angular momentum operator 
\begin{equation}\label{(angmom)}
\vec{L}\,=\,\vec{x}\times\vec{P}-\mu\frac{\vec{x}}{r}P_{5}\,.
\end{equation} 
These generators  satisfy the canonical commutation rules 
among themselves and with the components of all the other vector operators 
(e.g. coordinates,  momenta, etc.).

\section{The Dirac Field}
\

Let us denote by  $e(x)$ the  pentad fields that define the local frames and by 
$\hat e(x)$ those of the corresponding coframes. Their components, which give 
us the 
1-forms $\hat e^{\hat\alpha}=\hat e^{\hat\alpha}_{\mu}dx^{\mu}$ and 
the local derivatives  
$\hat\partial_{\hat\nu}=e^{\mu}_{\hat\nu}\partial_{\mu}$,  
have the usual orthonormalization properties 
$g_{\alpha\beta} e_{\hat\mu}^{\alpha}\, e_{\hat\nu}^{\beta}=
\eta_{\hat\mu \hat\nu},\, 
g^{\alpha\beta} \hat e^{\hat\mu}_{\alpha}\, \hat e^{\hat\nu}_{\beta}=
\eta_{\hat\mu \hat\nu},\, 
\hat e^{\hat\mu}_{\alpha} e^{\alpha}_{\hat\nu}=\delta^{\hat\mu}_{\hat\nu}$, 
where $\eta={\rm diag}(1,-1,-1,-1,-1)$ is the five-dimensional flat metric. 
In the case of the Taub-NUT geometry it is convenient to choose pentad 
fields with the following non-vanishing components \cite{P}
\begin{eqnarray}
&&\hat e^{0}_{0}=1\,, \quad
\hat e^{i}_{j}=\frac{1}{\sqrt{V}}\delta_{ij}\,, \quad
\hat e^{5}_{i}=\sqrt{V}A_{i}\,, \quad
\hat e^{5}_{5}=\sqrt{V}\,, \label{he}\\
&&e^{0}_{0}=1\,,\quad
e^{i}_{j}=\sqrt{V}\delta_{ij}\,,\quad
e^{5}_{i}=-\sqrt{V}A_{i}\,,\quad
e^{5}_{5}=\frac{1}{\sqrt{V}}\,.\label{e}
\end{eqnarray}
The components of the metric tensor, which can be written as  
$g_{\mu\nu}=\eta_{\hat\alpha \hat\beta}\hat e^{\hat\alpha}_{\mu}
\hat e^{\hat\beta}_{\nu}$ and 
$g^{\mu\nu}=\eta^{\hat\alpha \hat\beta} e_{\hat\alpha}^{\mu}
e_{\hat\beta}^{\nu}$,  raise or lower only the Greek 
indices (ranging from 0 to 5) while for the hat Greek ones 
(with the same range) we have to use the flat metric $\eta$. 
The commutation relations of the derivatives  
$\hat\partial_{\hat\nu}$ define the Cartan coefficients  
$C^{\,\cdot\cdot\,\hat\sigma}_{\hat\mu\hat\nu\cdot}=
e_{\hat\mu}^{\alpha} e_{\hat\nu}^{\beta}(\hat e^{\hat\sigma}_{\alpha,\beta}-
\hat e^{\hat\sigma}_{\beta,\alpha})$,
which will help us to write the spin connection in the local frames. 
   
The gauge group of the metric $\eta$ is $G(\eta)=SO(4,1)$. Its universal 
covering group, denoted by $\tilde G(\eta)$, is a subgroup of the $SU(2,2)$   
group which has a fundamental representation just in the space of the 
four-dimensional Dirac spinors. The five matrices 
 $\tilde\gamma^{\hat\alpha}$ which must satisfy
\begin{equation} 
\left\{ \tilde\gamma^{\hat\alpha},\, \tilde\gamma^{\hat\beta} \right\}
=2\eta^{\hat\alpha \hat\beta}, 
\end{equation}
can be defined in terms of standard Dirac matrices \cite{TH,DKK} as 
$\tilde \gamma^{0}=\gamma^{0},\,      
\tilde \gamma^{i}=\gamma^{i}$, ($i=1,2,3$),      
$\tilde \gamma^{5}=i\epsilon\gamma^{5}$ where we have introduced the parameter 
$\epsilon=\pm1$.
This choice has the advantage that all these five matrices are self-adjoint 
with respect to the usual Dirac conjugation, 
$\overline{\tilde\gamma^{\hat\alpha}}=
\gamma^{0}(\tilde\gamma^{\hat\alpha})^{\dagger}\gamma^{0}
=\tilde\gamma^{\hat\alpha}$.
Moreover, if we denote by  $S^{\hat\alpha \hat\beta}$ the covariant basis 
generators of the group $\tilde G(\eta)$ then we have 
\begin{eqnarray}
\left[\tilde\gamma^{\hat\alpha},\, \tilde\gamma^{\hat\beta} \right]
&=& -4iS^{\hat\alpha \hat\beta}\,,\\ 
\left[ S^{\hat\alpha \hat\beta},\, \tilde\gamma^{\hat\mu} \right] &=&
i(\eta^{\hat\beta \hat\mu}\tilde\gamma^{\hat\alpha}-
\eta^{\hat\alpha \hat\mu}\tilde\gamma^{\hat\beta})\,.
\end{eqnarray}

Let $\psi$ be a Dirac free field of  mass $M$, defined on the  space 
domain $D\times D_{5}$. Its gauge-invariant action  \cite{DKK} has the form     
\begin{equation}\label{(action)}
{\cal S}[\psi]=\int\, d^{5}x\sqrt{g}\left\{
\frac{i}{2}\,\left[\overline{\psi}\tilde\gamma^{\hat\mu}
\tilde\nabla_{\hat\mu}\psi-
(\overline{\tilde\nabla_{\hat\mu}\psi})\tilde\gamma^{\hat\mu}\psi
\right] - 
 M\overline{\psi}\psi\right\}
\end{equation}
involving the spin covariant derivatives 
$\tilde\nabla_{\hat\mu}=\hat\partial_{\hat\mu}+\Gamma_{\hat\mu}$ whose  
spin connection matrices are 
\begin{equation}\label{con}
\Gamma_{\hat\mu}
=\frac{i}{4}S^{\hat\nu \hat\lambda}(C_{\hat\mu \hat\nu \hat\lambda}+
C_{\hat\lambda \hat\mu \hat\nu}+C_{\hat\lambda \hat\nu \hat\mu})\,.
\end{equation}
Starting with  Eqs.(\ref{he}) and  (\ref{e}), after a little calculation,
we find  
\begin{equation}\label{sder}
\tilde \nabla_{i}=i\sqrt{V}P_{i}+\frac{i}{2}V\sqrt{V}\varepsilon_{ijk}
\Sigma_{j}^{*}B_{k}\,,\quad
\tilde \nabla_{5}=\frac{i}{\sqrt{V}}P_{5}-\frac{i}{2}V\sqrt{V}
\vec{\Sigma}^{*}\cdot\vec{B}\,,
\end{equation}
where
\begin{equation}
\Sigma_{i}^{*}=S_{i}+\frac{\epsilon}{2}\gamma^{5}\gamma^{i}\,,\quad
S_{i}=\frac{1}{2}\varepsilon_{ijk}S^{jk}\,. 
\end{equation}
Finally we obtain the Dirac equation, ${\cal D}\psi=0$, 
given by the Dirac operator  \cite{DIRAC} 
\begin{eqnarray}
{\cal D}&=&i\tilde\gamma^{\hat\alpha}\tilde\nabla_{\hat\alpha} - 
M\nonumber=i\gamma^{0}\partial_{t}-{\cal D}_{s}\\ 
&=&i\gamma^{0}\partial_{t} - \sqrt{V}\vec{\gamma}\cdot\vec{P}
-\frac{i\epsilon}{\sqrt{V}}\gamma^{5}P_{5}
-\frac{i\epsilon}{2} V\sqrt{V}\gamma^{5}\vec{\Sigma}^{*}\cdot\vec{B}-
M\,,\label{(de)}
\end{eqnarray}
which has the usual time-dependent term \cite{TH} and a {\em static} part, 
${\cal D}_{s}$, determined by the Taub-NUT geometry. 
The parameter values $\epsilon=\pm 1$ define two distinctive  versions of the 
theory related between themselves through
\begin{equation}\label{chir}
{\cal D}(-\epsilon, M)=-\gamma^{5} 
{\cal D}(\epsilon, -M)\gamma^{5}\,. 
\end{equation}
Hereby it results that these can be considered as chiral transformations to 
each other {\em only} when $M=0$.  We observe that in both cases the matrices 
$\Sigma_{i}^{*}$ are singular such that 
$\Sigma_{i}^{*}(\epsilon)+\Sigma_{i}^{*}(-\epsilon)=2S_{i}$. This is the motive 
why these versions appear as having quite different formalisms even if, in fact, 
they are equivalent as it results from Eq.(\ref{chir}). In order to do not 
produce confusion, in the following we work only in the version with 
$\epsilon=1$. We note that the Dirac operator of Ref.\cite{CH} is the static 
operator ${\cal D}_{s}$ of the version with $\epsilon=-1$ (and $M=0$).

By definition \cite{TH}, the Hamiltonian operator is 
$H=\gamma^{0}{\cal D}_{s}$. 
Other important observables are the generators 
of the global symmetry, $P_{5}$ and the  whole angular momentum operator 
$\vec{J}\,=\,\vec{L}+\vec{S}$. One can verify that its components, $J_{i}$, 
as well as $P_{5}$ are conserved in the sense that they commute with 
${\cal D}$ and $H$. This means that, as was expected, the Dirac equation is 
covariant under the transformations of the group $G_{s}$.  

\section{The regular Dirac equation}
\

The Dirac equation can be studied in the original above representation 
(with fixed $\epsilon=1$) or 
in any other representation obtained through a transformation, $\psi\to \psi'=
T\psi$, which can be unitary or not. Thus one can find suitable representations 
where the  behavior of the main operators under Hermitian conjugation or the 
supersymmetry of the Hamiltonian operator should be better pointed out. 

In our case it is convenient to write the Dirac equation in {\em regular} form 
with the help of the transformation $\psi_{R} =V^{-1/4} \psi$ \cite{DIRAC}. 
Using the identity \cite{CH}
\begin{equation}\label{idCH}
V^{\alpha}P_{i}V^{-\alpha}=P_{i}+i\alpha V B_{i}\,,
\end{equation}
we obtain the regular Dirac equation  ${\cal D}_{R}\psi_{R}=0$ given by  
the operator   
\begin{equation}\label{(rde)}
{\cal D}_{R}=i\gamma^{0}\partial_{t} - \sqrt{V}\vec{\gamma}\cdot\vec{P}
-\frac{i}{\sqrt{V}}\gamma^{5}P_{5}-
\frac{i}{2} V\sqrt{V}\gamma^{5}\vec{S}\cdot\vec{B}- M\,.
\end{equation}
In this form the Dirac equation  has usual terms apart  the specific 
Kaluza-Klein one and the magnetic term which 
couples the spin with the magnetic field like in the Schr\" odinger-Pauli 
non-relativistic theory. 

From the general theory of the Dirac field \cite{BD,DKK} 
we can deduce that the time-independent relativistic scalar product of two 
regular Dirac spinors is 
\begin{equation}\label{(sp)}
(\psi_{R},\psi_{R}')=\int_{D}\frac{d^{3}x}{\sqrt{V}}\int_{D_{5}}dx^{5}\,
\overline{\psi}_{R}(x)\gamma^{0}\psi_{R}'(x)\,. 
\end{equation}
Now we see that the regular form has the advantage of manifestly pointing out 
that the operators $\sqrt{V}P_{i}$ are 
self-adjoint with respect to the scalar product (\ref{(sp)}). Moreover, it is 
not difficult to convince ourselves that  $J_{i}$ and $P_{5}$ are also 
self-adjoint operators. 

The regular Hamiltonian 
operator, $H_{R}$, is given by ${\cal D}_{R}=\gamma^{0}(i\partial_{t}- 
H_{R})$.  
With the standard form of the Dirac matrices \cite{TH},
\begin{equation}
\gamma^{0}=\left(
\begin{array}{cc}
1&0\\
0&-1
\end{array}
\right)\,,\quad
\gamma^{i}=\left(
\begin{array}{cc}
0&\sigma_{i}\\
-\sigma_{i}&0
\end{array}
\right)\,,\quad
\gamma^{5}=\left(
\begin{array}{cc}
0&1\\
1&0
\end{array}
\right)\,,
\end{equation}
(where $\sigma_{i}$ are the Pauli matrices), we obtain 
\begin{equation}
H_{R}=\left(
\begin{array}{cc}
M&\Pi^{\dagger}\\
\Pi&-M
\end{array}\right)
\end{equation}
where 
\begin{eqnarray}\label{pili}
\Pi&=&\sqrt{V}\vec{\sigma}\cdot\vec{P}-\frac{i}{\sqrt{V}}P_{5}-
\frac{i}{4}V\sqrt{V}\vec{\sigma}\cdot\vec{B}\nonumber\\
&=&V^{-1/4}\left(\sqrt{V}\vec{\sigma}\cdot\vec{P}-\frac{i}
{\sqrt{V}}P_{5} \right)V^{1/4}\,,
\end{eqnarray}
as it results from  Eq.(\ref{idCH}). 

In this representation it is obvious that $H_{R}$ is self-adjoint  
and has {\em manifest} supersymmetry. 
This means that the two-component spinors of 
$\psi_{R}=(\psi_{R}^{(1)},\,\psi_{R}^{(2)})^{T}$ satisfy 
the equations with second order derivatives,
\begin{equation}
(\Pi^{\dagger}\,\Pi +M^2+{\partial_{t}}^{2})\psi_{R}^{(1)}=0\,,\quad
(\Pi\,\Pi^{\dagger} +M^2+{\partial_{t}}^{2})\psi_{R}^{(2)}=0\,,\quad
\end{equation}
that can be obtained directly from $({H_{R}}^2 +{\partial_{t}}^{2})\psi_{R}=0$.

In the original form of the Dirac equation these properties appear 
as partially hidden since the transformation which leads to the regular 
equation is not unitary. Other interesting equivalent forms of the Dirac 
equation can be obtained by using also non-unitary transformations. For example 
the transformation $T={\rm diag}(1,\, V^{-1/2})$ leads to a representation 
in which the magnetic term is completely absorbed. However, what is important 
is that there is at least one representation, namely the regular one, where 
we find  good  Hermitian properties and the manifest supersymmetry of the 
Hamiltonian.

\section{The energy eigenspinors}
\

Bearing in mind that the energy, $E$, is conserved it is convenient to 
consider particular solutions with fixed energy that can be of positive or 
negative frequency like in the case of the Dirac theory in flat spacetime. 
We denote them by
\begin{equation}
\psi_{E}^{(+)}(x)=e^{-iEt}u_{E}(\vec{x},x^{5})\,,\quad  
\psi_{E}^{(-)}(x)=e^{iEt}v_{E}(\vec{x},x^{5})\,,  
\end{equation}
and say that $u_{E}$ is the particle energy eigenspinor while $v_{E}$ is 
the antiparticle one since they satisfy $(H-E)u_{E}=0$ and $(H+E)v_{E}=0$.   

Let us analyze the particle-like eigenvalue problems  starting 
with the Hamiltonian operator put in the form 
\begin{equation}\label{HH}
H =\left(
\begin{array}{cc}
M&V\pi^{*}\frac{\textstyle 1}{\textstyle \sqrt{V}}\\
\sqrt{V}\pi&-M
\end{array}\right)
\end{equation}
where we have denoted by
\begin{equation}
\pi=\,\vec{\sigma}\cdot\vec{P}-\frac{iP_{5}}{V}\,,\quad 
\pi^{*}=\,\vec{\sigma}\cdot\vec{P}+\frac{iP_{5}}{V}\,,\quad 
\end{equation}
the operators which give 
\begin{equation}\label{dpipi}
\Delta=V\,\pi^{*}\pi\,.
\end{equation}
Then we find that the  two-component spinors of 
$u_{E}=(u^{(1)}_{E},\,u^{(2)}_{E})^{T}$ satisfy
the following  equations
\begin{eqnarray}
V\pi^{*}\frac{1}{\sqrt{V}}\,u^{(2)}&=&(E-M)\,u^{(1)}\,,\label{E10}\\
\sqrt{V}\pi\,u^{(1)}&=&(E+M)\,u^{(2)}\,,\label{E20}
\end{eqnarray}
which, according to Eq.(\ref{dpipi}), are equivalent with
\begin{eqnarray}
\Delta\, u_{E}^{(1)}&=&(E^{2}-M^{2})\,u_{E}^{(1)}\,,\label{E1}\\
u^{(2)}_{E}&=&\frac{1}{E+M}\sqrt{V}\,\pi\,u^{(1)}_{E}\,.\label{E2}
\end{eqnarray}
This  remarkable result shows us that $u_{E}^{(1)}$ may be {\em any} solution 
of the scalar equation (\ref{E1}) which is nothing else than the static 
Klein-Gordon equation in  Taub-NUT geometry. In other respects,   
it is clear that $u^{(2)}_{E}$ is completely determined by Eq.(\ref{E2}) 
if the form of $u^{(1)}_{E}$ is given.

Hence the  conclusion is that the Dirac equation in  Taub-NUT geometry  
is always analytically solvable since the Klein-Gordon equation (\ref{E1}) is 
solvable having similar solutions as the Schr\" odinger one \cite{CV}. The 
selection of the solutions can be done  starting with desired  
spinors $u^{(1)}_{E}$ that satisfy Eq.(\ref{E1}) and then calculating 
$u^{(2)}_{E}$. Since the Klein-Gordon equation is scalar, the form of the 
spinor  $u^{(1)}_{E}$ and, therefore, that of $u_{E}$ is strongly dependent on 
the choice of the spin observables included in the complete set of the 
commuting operators which defines the quantum modes.

\section{Observables}
\

The knowledge of the general form of the solutions  offers us the 
opportunity to analyze the structure and  properties of the 
conserved observables. These are operators in the spaces of the spinors 
$u_{E}$ which can be expressed in terms  of operators acting upon the 
two-component spinors \cite{DYON}, $\pi,\,\pi^{*},\,_{2}J_{i}=
L_{i}+\sigma_{i}/2$ and 
\begin{equation}
\sigma_{r}=\frac{1}{r}\vec{\sigma}\cdot\vec{x}\,,\quad
\sigma_{P}=\,\vec{\sigma}\cdot\vec{P}\,,\quad
\sigma_{L}=\,\vec{\sigma}\cdot\vec{L}\,.
\end{equation}
The problem we briefly investigate here is how may look the 
operators which commute with the  Hamiltonian (\ref{HH}). 

The simplest operators have the diagonal form 
\begin{equation}
X=\left(
\begin{array}{cc}
X^{(1)}&0\\
0&X^{(2)}
\end{array}\right)
\end{equation}
where $X^{(1)}$ and $X^{(2)}$ are $2\times 2$ matrix operators.
The condition $[X,\,H]=0$ is equivalent with
\begin{equation}\label{x12}
X^{(2)}\sqrt{V}\pi=\sqrt{V} \pi X^{(1)}
\,,\quad V\pi^{*}\frac{1}{\sqrt{V}}X^{(2)}=
X^{(1)}V\pi^{*}\frac{1}{\sqrt{V}}\,,
\end{equation}
from which it results that $[X^{(1)},\,\Delta]=0$. Therefore, we can start 
with an operator $X^{(1)}$ which commutes with $\Delta$ and to try to find 
the operator $X^{(2)}$ which satisfies Eqs.(\ref{x12}). If this problem 
has solution and  we have
\begin{equation}\label{Xxiu}
X^{(1)}u^{(1)}_{E}=\xi\, u_{E}^{(1)}\,,
\end{equation}
then the energy eigenspinor $u_{E}$ is an eigenspinor of  
$X$  corresponding to the same eigenvalue, $\xi$. The best example is 
represented by the components of the whole angular momentum, 
$J_{i}={\rm diag}(_{2}J_{i},\,_{2}J_{i})$, which commute with 
$H$ since $_{2}J_{i}$ commute with $\pi$, $\pi^{*}$ and $V$.     

If we have an operator $X^{(1)}$ which commutes with $\Delta$ and 
satisfies Eq.(\ref{Xxiu}) but there is no satisfactory solution for $X^{(2)}$ 
we can introduce  an observable starting with the singular operator 
$X_{0}={\rm diag}(X^{(1)},\,0)$ that commutes with $H^2$. This property 
guarantees that the new operator
\begin{equation}\label{Qx}
{\cal Q}(X^{(1)})=\left\{X_{0},\,H \right\}\,, 
\end{equation}
commutes with $H$. In addition, by using Eqs.(\ref{E10}), (\ref{E20}) and 
(\ref{Xxiu}) we obtain the interesting eigenvalue equation 
\begin{equation}\label{lugu}
{\cal Q}(X^{(1)}) u_{E}= (E+M)\,\xi\,u_{E}\,. 
\end{equation} 
Thus we arrive to another important result namely that for any operator 
which commutes with $\Delta$ there exists an operator in the space of the Dirac 
energy eigenspinors $u_{E}$ of the form (\ref{Qx}) which commutes with 
$H$. Particularly, for $X^{(1)}=1$  we get ${\cal Q}(1)=H+M$.  
On the other hand, the mapping $X^{(1)}\to {\cal Q}(X^{(1)})$ cannot be 
interpreted as a representation since, in general, for two operators 
$X^{(1)}$ and $Y^{(1)}$ which commute with $\Delta$, we have 
$[ {\cal Q}(X^{(1)}),\,  {\cal Q}(Y^{(1)})]\not= 
{\cal Q}([X^{(1)},\,Y^{(1)}])$. 
Fortunately, if $X^{(1)}$ and $Y^{(1)}$ commute with each other then it is true 
that $[ {\cal Q}(X^{(1)}),\,  {\cal Q}(Y^{(1)})]=0$. 

However, apart from the above cases, there are many other types of conserved 
operators constructed by using algebraic operations or directly related to 
the specific geometric objects of the Taub-NUT geometry, as St\" ackel-Killing 
and Killing-Yano tensors. In the next Section we shall present some 
operators connected with the covariantly constant 
Killing-Yano tensors of the Taub-NUT geometry.

\section{Spin-like and Dirac-type operators}
\

As observed in \cite{GR}, the Taub-NUT geometry possesses four 
Killing-Yano tensor of valence $2$:
\begin{equation}\label{(yano)}
f_{\mu(\nu;\lambda)} = 0
\end{equation}
where the round bracket denotes symmetrization over the indices enclosed. 
The first three, 
\begin{eqnarray}\label{KY}
f^i 
&=& f^i_{{\hat \alpha}{\hat \beta}} {\hat e}^{\hat \alpha} \wedge 
{\hat e}^{\hat \beta}\nonumber\\
&=& 2\mu (dx^5 + \vec{A}\cdot d\vec{x})\wedge dx^i +\varepsilon_{ijk} V^{-1} 
dx^j\wedge dx^k\nonumber\\
&=& 2 {\hat e}^5\wedge  {\hat e}^i +\varepsilon_{ijk} {\hat e}^j\wedge 
{\hat e}^k\,,
\end{eqnarray}
are rather special since they are covariantly constant (with 
vanishing field strength). 

Using these  tensors  we can construct the following spin-like operators,
\begin{equation}\label{sl}
\Sigma_{i}=\frac{i}{4} f^{i}_{\hat\alpha\hat\beta}\gamma^{\hat\alpha}
\gamma^{\hat\beta}=S_{i}-\frac{1}{2}\gamma^{5}\gamma^{i}=
\left( \begin{array}{cc}
\sigma_{i}&0\\
0&0
\end{array}\right) \,,
\end{equation}
which have similar properties to that of the Pauli 
matrices $\sigma_i$ and, in addition, satisfy
\begin{equation}\label{Si}
\Sigma_{i}+\Sigma^{*}_{i}=2S_{i}\,.
\end{equation}
Since the Pauli matrices commute with $\Delta$, the spin-like operators 
(\ref{sl}) commute with $H^2$ and, therefore, we can introduce the 
operators of the form (\ref{Qx})  
\begin{equation}\label{Qi}
{\cal Q}(\sigma_{i})=\{\Sigma_{i},\,H\}=
\left( \begin{array}{cc}
2M\sigma_{i}&V\sigma_{i}\pi^{*}\frac{\textstyle 1}{\textstyle \sqrt{V}}\\
\sqrt{V}\pi\sigma_{i}&0
~\end{array}\right)
\end{equation}
which commute with the Hamiltonian  (\ref{HH}). Let us observe that they 
have the remarkable property:
\begin{equation}\label{QQ}
\{ {\cal Q}(\sigma_{i}),\,{\cal Q}(\sigma_{j})\}=2\delta_{ij}(H+M)^2\,.
\end{equation}  

The existence of the Killing-Yano tensors (\ref{KY}) allows one to 
construct generalizations of the operator ${\cal D}_{s}$ defined by
\begin{equation}
{\cal D}_{i}=f^{i}_{\hat\alpha\hat\beta}\tilde\gamma^{\hat\alpha}\tilde\nabla
^{\hat\beta}\,,
\end{equation}
and called Dirac-type operators \cite{G3}. After a little calculation we find 
that their explicit form is
\begin{equation}\label{Dt}
{\cal D}_{i}=
\left( \begin{array}{cc}
0&V\sigma_{i}\pi^{*}\frac{\textstyle 1}{\textstyle \sqrt{V}}\\
\sqrt{V}\pi\sigma_{i}&0
\end{array}\right)
\end{equation}
and from Eqs.(\ref{sl}), (\ref{Qi}) and (\ref{Dt}) we get
\begin{equation}
{\cal Q}(\sigma_{i})=2M\Sigma_{i}+{\cal D}_{i}\,.
\end{equation}
In addition, we have 
\begin{equation}\label{cocofix}
\{\gamma^{0},\,{\cal D}_{i}\}=0\,,
\end{equation}
and
\begin{equation}
[H,\,\Sigma^{i}]=-\gamma^{0}{\cal D}_{i}\,,\quad
[H,\,{\cal D}_{i}]=2M\gamma^{0}{\cal D}_{i}\,.
\end{equation}

For $M = 0$, the  operators  ${\cal Q}(\sigma_{i})$ coincide 
with the Dirac-type operators ${\cal D}_{i}$. Moreover, including into 
Eq.(\ref{QQ}) the static part of ${\cal D}$, denoted now by $Q_{0}=
i{\cal D}_{s}=i\gamma^{0}H$,
by the side of the operators $Q_{i}={\cal Q}(\sigma_{i})$ we obtain 
\begin{equation}\label{QQH}
\{Q_{A},\,Q_{B}\}=2\delta_{AB}H^2\,, \quad A,B,...=0,1,2,3\,.
\end{equation}  
It is worthy of notice that these relations correspond to the $N = 4$ 
supersymmetry algebra realized by the supercharges of the spinning 
Taub-NUT model \cite{GR,G2,G3}. In fact  relations of the form (\ref{QQH})
make manifest the link between the existence of the covariantly constant
Killing-Yano tensors (\ref{KY}) and the hyper-K\" ahler geometry of the 
Taub-NUT manifold. In other respects, let us observe that Eq.(\ref{cocofix})
permits to convert the anticommutator (\ref{QQH}) into a 
commutation relation between the operators $\gamma^0 Q_A$ and $Q_B$ for 
$A\neq B$.

The fourth Killing-Yano tensor of the Taub-NUT space is not covariantly 
constant. This tensor is involved in the construction of 
a conserved vector analogous to the Runge-Lenz vector of the Kepler problem
\cite{GR,G2,G3}.
The existence of the extra conserved quantities, quadratic in velocities, 
implies the possibility of separating variables for Dirac equation in two
different coordinate systems. This matter will be discussed elsewhere 
\cite{CV1}. 

\section{Central modes}
\

In this section we shall show in what manner the central modes can be 
defined using  an appropriate complete set of commuting observables. 
We  consider the local chart with spherical coordinates,  $r, \theta, \phi$, 
commonly related to the Cartesian ones  and the new coordinate $\chi$ 
defined as 
\begin{equation}
\chi=-\frac{1}{\mu}x^{5}-{\rm arctan}\frac{x^2}{x^1}\,.
\end{equation}
Then the line element reads 
\begin{equation}
ds^{2}=dt^{2}-\frac{1}{V}(dr^{2}+r^{2}d\theta^{2}+ 
r^{2}\sin^{2}\theta\, d\phi^{2})-\mu^{2}V(d\chi+\cos\theta\, d\phi)^{2}\,,  
\end{equation}
since $A_{r}=A_{\theta}=0$ and $A_{\phi}=\mu(1-\cos\theta)$. Note that in 
this chart   $r\in D_{r}=\{r|V(r)>0\}$ (i.e., $r>0$ if $\mu>0$ or $r>|\mu|$ if 
$\mu<0$), the angular coordinates $\theta,\,\phi$ cover the sphere $S^{2}$ 
while $\chi\in D_{\chi}=[0,4\pi)$.

In the following it is convenient to use the orbital operators in spherical 
coordinates \cite{CV} and  $Q=-\mu P_{5}=-i\partial_{\chi}$ instead of $P_{5}$. 
We assume that the central modes of the Dirac field in Taub-NUT geometry are 
given by the common eigenspinors of the {\em complete} set of commuting 
operators 
$\{H,\,Q,\,{\vec{J}}^{2},\,J_{3},\,{\cal Q}(\sigma_{L}+1)\}$.     
These eigenspinors  have the form
\begin{equation}\label{spinor}
u^{\pm}_{E,q,j,m_{j}}
=N^{\pm}_{E,q,j}\frac{1}{r}\left(
\begin{array}{r}
-i(E+M)\,f^{\pm}_{E,q,j}\,\Psi^{\pm}_{q,j,m_{j}}\\
\sqrt{V}(h^{\pm} _{E,q,j}\,\Psi^{\pm}_{q,j,m_{j}}+
g^{\pm} _{E,q,j}\,\Psi^{\mp}_{q,j,m_{j}})\,,
\end{array}\right)
\end{equation}
where $N^{\pm}$ are normalization factors, the radial functions $f^{\pm},\,
g^{\pm}$ and $h^{\pm}$ depend only on $r$ while $\Psi^{\pm}$ are the 
spherical spinors defined in Appendix. These completely solve the common 
eigenvalue problems of the operators 
$Q,\,{\vec{J}}^{2}$,  $J_{3}$ and ${\cal Q}(\sigma_{L}+1)$ for the eigenvalues 
$q,\,j(j+1)$, $m_{j}$, and $\pm(E+M)(j+{1\over 2})$, respectively. 

Then it remains only to solve  the energy eigenvalue problem
which must give the radial functions  
according to Eqs.(\ref{E1}) and (\ref{E2}). We find first that $f^{\pm}$ 
is a solution of the radial Klein-Gordon equation,
\begin{equation}\label{(kgrad)}
\left[-\frac{d^2}{dr^2}+\frac{l_{\mp}(l_{\mp}+1)}{r^2}-\frac{\alpha}{r}\right]
f^{\pm}_{E,q,j}(r)=\beta f^{\pm}_{E,q,j}(r)\,, 
\end{equation}
which has the {\em same} form as the Schr\" odinger one \cite{CV} but with the 
parameters, $l_{\pm}=j\pm \frac{1}{2}$ and
\begin{equation}\label{(ab)}
\alpha=\mu\left[E^{2}-M^{2}-2\frac{q^2}{\mu^2}\right]\,,\quad
\beta=E^{2}-M^{2}-\frac{q^2}{\mu^2}\,.
\end{equation}
The other radial functions can be calculated from Eq.(\ref{E2}) if we take 
into account that
\begin{equation}
i\sigma_{P}=\sigma_{r}\left[\partial_{r}+\frac{1}{r}-\frac{\sigma_{L}+1}{r}
\right]+\frac{Q}{r}
\end{equation}
and by using Eqs.(\ref{kkk}) and (\ref{pmmp}). We obtain the radial 
equations 
\begin{eqnarray}
g^{\pm}_{E,q,j}&=&\sqrt{1-(\lambda^{q}_{j})^{2}}\left(-\frac{d}{dr}
\pm\frac{j+\frac{1}{2}}{r}\right)f^{\pm}_{E,q,j}\,,\\  
h^{\pm}_{E,q,j}&=&\lambda^{q}_{j}\left(\mp\frac{d}{dr}
+\frac{j+\frac{1}{2}}{\mu V}\right)f^{\pm}_{E,q,j}\,,
\end{eqnarray}
which lead to the identity
\begin{equation}
h^{\pm}_{E,q,j}=\frac{q}{\mu}\,f^{\pm}_{E,q,j}\pm 
\frac{\lambda^{q}_{j}}{\sqrt{1-(\lambda^{q}_{j})^2}}\,g^{\pm}_{E,q,j}\,.  
\end{equation}
The last step is to calculate the normalization constants with the help  
of the scalar product (\ref{(sp)}) where the spinors $\psi_{R}$ 
will be replaced by $V^{-1/4}u^{\pm}_{E,q,j,m_{j}}$. Since the spherical 
spinors are orthogonal, we obtain for the eigenspinors corresponding to 
the discrete energy levels that
\begin{equation}\label{(scprod1)}
\frac{1}{|N^{\pm}_{E,q,j}|^2}=(E+M)^{2}\int_{D_{r}}\frac{dr}{V} 
|f^{\pm}_{E,q,j}|^{2} +\int_{D_{r}}dr \,\left(
|g^{\pm}_{E,q,j}|^{2} +
|h^{\pm}_{E,q,j}|^{2} \right)\,.
\end{equation}

The radial Klein-Gordon equation (\ref{(kgrad)}) produces the same central 
modes as those discussed in Ref.\cite{CV} with similar energy spectra. 
Therefore the energy levels are deeply degenerated having no fine structure. 
For example, the discrete levels of the case $\mu<0$, 
\begin{equation}\label{(een)}
E_{n}^{2}=M^{2}+\frac{2}{\mu^2}\left[n\sqrt{n^{2}-q^{2}}-(n^{2}-q^{2})
\right]\,,
\end{equation}        
depend only on the principal quantum number $n$ which takes all the integer 
values allowed by the selection rule $|q|<j+\frac{1}{2}<n$ derived from those 
presented in Appendix. It is interesting to observe that the condition $n>|q|$ 
eliminates the zero modes (with $E=M$) for $q\not =0$. Like in the scalar case 
the energy spectrum is countable with finite range since 
\begin{equation}
\lim_{n\to \infty} E_{n}^{2}={M_{ef}}^{2}=M^{2}+\frac{q^2}{\mu^2}\,,  
\end{equation}
where $M_{ef}$ play the role of the effective mass of the spin half particle 
in the field of the Dirac monopole. This arises from the usual mass term of 
${\cal D}$ combined with the specific Kaluza-Klein contribution 
\cite{DKK,DIRAC}. We must specify that in the case of $M=0$ this spectrum is 
similar with that of Ref.\cite{CH} but the energy eigenspinors as well as the 
selection rules of the quantum numbers are different.   

Finally we observe that there are more other conserved observables which 
commute with those of the complete set that defines the central modes. Indeed, 
if we take the operators $\vec{L}^{2}$, $_{2}\vec{J}^{2}$ and $_{2}J_{3}$ 
which commute with $\Delta$,  we can verify that the operators 
${\cal Q}(\vec{L}^{2})$, ${\cal Q}(_{2}\vec{J}^{2})$ and ${\cal Q}(_{2}J_{3})$ 
commute among themselves and with all the other observables considered above.  
Consequently, the eigenspinors (\ref{spinor}) of the central basis are, in 
addition, common  eigenspinors of these new  observables, corresponding to the 
eigenvalues   $(E+M)l_{\mp}(l_{\mp}+1)$, $(E+M)j(j+1)$ and $(E+M)m_{j}$, 
respectively.

\section{Conclusions}
\

The main conclusion of this article is that the $SO(4,1)$ gauge-invariant 
theory of the Dirac field in Taub-NUT geometry leads to an analytically 
solvable field equation which gives similar energy levels like those of the 
scalar modes. In general, these are deeply degenerated having no fine structure. 
For this reason we need to use many commuting operators in order to completely 
determine the quantum modes of the Dirac field. Fortunately, in this theory we 
have to handle a collection of conserved observables much more larger than that 
of the scalar field. This is because one can associate to each operator which 
commutes with the Klein-Gordon operator at least one operator which should 
commute with the Dirac Hamiltonian. In this way we have found the new operator 
${\cal Q}(\sigma_{L}+1)$ necessary to complete the  set of commuting operators  
$\{H,\,Q,\,{\vec{J}}^{2},\,J_{3}\}$  and the other new conserved observables 
that are diagonal in the basis of the eigenspinors (\ref{spinor}). However, as 
mentioned, there is the possibility that  other kind of conserved observables 
could be constructed using the specific Killing-Yano tensors of the Taub-NUT 
geometry.      

In a forthcoming paper \cite{CV1} we shall analyze the consequences of the
presence of a conserved vector analogous to the Runge-Lenz vector of the
Kepler-type problem. The existence of this conserved quantity is rather
surprising in view of the complexity of the equations of geodesic motion
in the Taub-NUT space. The components of the Runge-Lenz type vector are
Killing tensors and they can be expressed as symmetrized products of the
covariantly constant Killing-Yano tensors $f^i$ (\ref{KY}) and a fourth
Killing-Yano tensor $f^Y$ non-covariantly constant \cite{GR,G3}. The first 
three tensors, $f^i$, transform as vector under spatial rotations generated by 
${\vec J}$, while the last one, $f^Y$, is a scalar. The existence of the
extra conserved quantities, quadratic in four-velocities, implies the
possibility of separating variables in two different coordinate systems.

Finally we intend to extend the study of the Dirac equations on
generalized Taub-NUT metrics \cite{IWK}. This class
of metrics admits a Kepler-type symmetry, but, in general, the Killing
tensors involved in the Runge-Lenz vector cannot be expressed as products
of Killing-Yano tensors \cite{MV}.
It is expected that the non-existence of the Killing-Yano
tensors makes the study of "hidden" symmetries more laborious for
relativistic particles with spin ${1\over 2}$.

\subsection*{Acknowledgments}
\

We should like to thank Marc Henneaux and Yavuz Nutku for interesting and
helpful conversations concerning the motion of fermions on curved spaces.

\appendix
\section{Spherical spinors}
\

We define the spherical spinors in  usual manner \cite{TH} starting with our 
$SO(3)\otimes U(1)$ harmonics, $Y^{q}_{l,m}$, introduced in Ref.\cite{CV}. 
These are more general than others \cite{HARM} since they are defined for  
any real value of $q$ without restrictions upon the values of $l$ and $m$ 
apart the selection rule $|q|-1<|m|\le l$. 

In the case of the Taub-NUT geometry the boundary conditions on 
$S^{2}\times D_{\chi}$ require $l$ and $m$ to be integer numbers while $q=0,\pm 
1/2,\pm 1,...$ \cite{G2}. However, for integer $l$ the spherical spinors 
can be defined for any  $q$ according to the usual method of the decomposition 
of the direct product of $SU(2)$  representations  by  taking the 
$SO(3)\otimes U(1)$ harmonics instead of the usual ones. Thus we find that the 
spherical spinors,  $\Psi^{\pm}_{q,j,m_{j}}(\theta,\phi,\chi)$, 
are the common two-component eigenspinors of the eigenvalue problems  
\begin{eqnarray}
Q\,\Psi^{\pm}_{q,j,m_{j}}&=& q\,\Psi^{\pm}_{q,j,m_{j}}\,, \\
{_2}\vec{J}^{2}\,\Psi^{\pm}_{q,j,m_{j}}&=& j(j+1)\,\Psi^{\pm}_{q,j,m_{j}}
\,, \\
{_{2}}J_{3}\,\Psi^{\pm}_{q,j,m_{j}}&=& m_{j}\,\Psi^{\pm}_{q,j,m_{j}}\,, \\
(\sigma_{L}+1)\,\Psi^{\pm}_{q,j,m_{j}}&=& \pm(j+1/2)\,
\Psi^{\pm}_{q,j,m_{j}}\,.\label{kkk}  
\end{eqnarray}
They are, in addition, eigenfunctions of ${\vec{L}}^{2}$ corresponding to the 
eigenvalues $l(l+1)$ with $l=j\pm\frac{1}{2}$. For $j=l+\frac{1}{2}
>|q|-\frac{1}{2}$ we have 
\cite{TH,CV}  
\begin{equation}
\Psi^{+}_{q,j,m_{j}}=\frac{1}{\sqrt{2j}}\left(
\begin{array}{l}
\sqrt{j+m_{j}}\, Y^{q}_{j-\frac{1}{2},m_{j}-\frac{1}{2}}\\
\sqrt{j-m_{j}}\, Y^{q}_{j-\frac{1}{2},m_{j}+\frac{1}{2}}
\end{array}\right)
\end{equation}
while for $j=l-\frac{1}{2}>|q|-\frac{3}{2}$ we get
\begin{equation}
\Psi^{-}_{q,j,m_{j}}
=\frac{1}{\sqrt{2j+2}}\left(
\begin{array}{l}
\sqrt{j-m_{j}+1}\, Y^{q}_{j+\frac{1}{2},m_{j}-\frac{1}{2}}\\
-\sqrt{j+m_{j}+1}\, Y^{q}_{j+\frac{1}{2},m_{j}+\frac{1}{2}}
\end{array}\right)\,.
\end{equation}
We specify that these spherical spinors  are orthonormal since the 
$SO(3)\otimes U(1)$ harmonics are orthonormal with respect to the 
angular scalar product \cite{CV}.  

Finally, using Eq.(\ref{kkk}) and the superalgebra
\begin{equation}
{\sigma_{r}}^{2}=1\,,\quad \{\sigma_{r},\,\sigma_{L}+1\}=2q
\end{equation}
we find that    
\begin{equation}\label{pmmp}
\sigma_{r}\Psi^{\pm}_{q,j,m_{j}}=
\pm\lambda^{q}_{j}\,\Psi^{\pm}_{q,j,m_{j}}
+\sqrt{1-(\lambda^{q}_{j})^{2}}\,\Psi^{\mp}_{q,j,m_{j}}
\end{equation}
where we have denoted $\lambda^{q}_{j}=q/(j+\frac{1}{2})$. Note that a 
similar property is reported in Ref.\cite{HARM1}.

\end{document}